\documentclass[twocolumn,amsmath,amssymb,nofootinbib]{revtex4}
\usepackage{graphicx}
\usepackage{dcolumn}
\usepackage{color}
\usepackage{bm}
\usepackage{amsmath,amssymb,graphicx}
\usepackage{epsfig}
\usepackage{enumerate}
\usepackage{array}
\usepackage{multirow}
\usepackage{tikz-cd}

\newtheorem{prop}{Proposition}

\begin{document}
\title
{Hamilton-Jacobi approach to thermodynamic transformations}
\author{Aritra Ghosh\footnote{ag34@iitbbs.ac.in}}
\affiliation{School of Basic Sciences,\\ Indian Institute of Technology Bhubaneswar, Argul, Jatni, Khurda, Odisha 752050, India}
\vskip-2.8cm
\date{\today}
\vskip-0.9cm

\textit{Dedicated to Dr Asok Kumar Das on the occasion of his 50th birthday with deep respect and admiration}

\vspace{5mm}
\begin{abstract}
In this note, we formulate and study a Hamilton-Jacobi approach for describing thermodynamic transformations. The thermodynamic phase space assumes the structure of a contact manifold with the points representing equilibrium states being restricted to certain submanifolds of this phase space. We demonstrate that Hamilton-Jacobi theory consistently describes thermodynamic transformations on the space of externally controllable parameters or equivalently, the space of equilibrium states. It turns out that in the Hamilton-Jacobi description, the choice of the principal function is not unique but, the resultant dynamical description for a given transformation remains the same irrespective of this choice. Some examples involving thermodynamic transformations of the ideal gas are discussed where the characteristic curves on the space of equilibrium states completely describe the dynamics. The geometric Hamilton-Jacobi formulation which has emerged recently is also discussed in the context of thermodynamics. 
\end{abstract}

\maketitle

\tableofcontents

\section{Introduction}
Hamilton-Jacobi theory is well known in classical mechanics as an alternate yet equivalent formulation to Hamiltonian dynamics. In Hamiltonian dynamics, one considers a phase space which is typically a symplectic manifold \(M\) with symplectic form \(\omega\) whose local (Darboux) coordinates \(\{q^i,p_i\}\) are identified with the coordinates and momenta of the mechanical system \cite{goldsteincm,arnoldcm}. For any Hamiltonian function \(H \in C^\infty(M)\) on this phase space, there exists a vector field \(X_H\) given by
\begin{equation}
	\iota_{X_H} \omega = dH.
\end{equation} Since \(\omega\) is invertible, the correspondence between \(dH\) and \(X_H\) is an isomorphism. Here, \(X_H\) is called the Hamiltonian vector field whose integral curves are the solutions to the standard Hamilton's equations of motion: 
\begin{equation}\label{SHeqns}
	\dot{q}^i = \frac{\partial H}{\partial p_i}, \hspace{5mm} \dot{p}_i = - \frac{\partial H}{\partial q^i}. 
\end{equation} In the context of a mechanical system, it is often useful to think of the phase space as being the cotangent bundle of the configuration space, i.e. \(M = T^*Q\). The generalized coordinates of the system \(\{q^i\}\) are coordinates on \(Q\) whereas the momenta \(\{p_i\}\) are the fiber coordinates. This idea of the configuration space can be generalized to the notion of Lagrangian submanifolds of \(M\) on which the local coordinates are \(\{q^i,p_j\}\) where  \(i \in I, j \in J\) with \(I \cup J = \{1, \cdots , n\}\) and \(I \cap J = \phi\). In other words, a Lagrangian submanifold cannot include a conjugate pair of variables. \\

In contrast to the Hamiltonian approach, the Hamilton-Jacobi problem is formulated on the configuration space itself (or more generally a Lagrangian submanifold). If \(Q\) be the configuration space of the system with local coordinates \(\{q^i\}\) where \( i \in \{1, \cdots , n\}\), then the Hamilton-Jacobi problem is to seek a function \(W = W(q^i,t)\) satisfying
\begin{equation}
 dW - p_i dq^i + H dt = 0.
\end{equation} In mechanics, \(W\) is often called the principal function. This leads to the following relations:
\begin{equation}
	p_i = \frac{\partial W(q^i,t)}{\partial q^i}, \hspace{5mm} \frac{\partial W(q^i,t)}{\partial t} = - H \Bigg( q^i,   \frac{\partial W(q^i,t)}{\partial q^i}, t \Bigg). 
\end{equation}
Note that \(W : Q \times \mathbb{R} \rightarrow \mathbb{R}\) where \(t \in \mathbb{R}\). Thus, the momenta have disappeared from the picture and they can be derived from the knowledge of the principal function. This formulation of describing the dynamics can be shown to be equivalent to Eqs. (\ref{SHeqns}) (see for example \cite{goldsteincm}). \\

Contact geometry \cite{Geiges,Arnold,CM1,CM2} is understood to be the proper geometric setting for thermodynamics \cite{RT1,RT2,RT3,RT4,RT5,RT6,RT7,RT8} (see also \cite{hermann,Peter79}). The thermodynamic phase space is equipped with the structure of a contact manifold on which the local coordinates are the different thermodynamic variables while various thermodynamic processes are understood as contact Hamiltonian flows. This description is quite appealing because it sets thermodynamics on a similar footing with geometric mechanics with several common features such as the existence of spaces of equilibrium states in the former in analogy with configuration spaces encountered in the latter. Furthermore, the thermodynamic geometry of Ruppeiner (motivated from fluctuation theory) \cite{Ruppeiner,Ruppeiner1} and Weinhold \cite{Weinhold} appears quite naturally in such a setting with clear connections to thermodynamic fluctuation theory and equilibrium statistical mechanics \cite{mstatistical}. These formalisms have been extended to more exotic systems such as black holes \cite{BH1,BH2} (see \cite{BH3,BH4} for some recent developments). A thorough formalism of extended black hole thermodynamics within the framework of contact geometry has been developed recently in \cite{contactBH}  (also see \cite{RT6,Baldiotti} for some older works) and various black holes in the anti-de Sitter spacetime have been shown to emerge from their ideal gas limit by using suitable contact vector fields which deform the ideal gas equation of state. \\

In the present work, we study thermodynamic transformations using a Hamilton-Jacobi approach. The principal function is defined on the spaces of equilibrium states with the Hamilton-Jacobi equation describing its time evolution. This approach is in contrast to the Hamiltonian approach adopted in the earlier studies \cite{RT4,RT5,contactBH} on thermodynamic transformations. Although both the methods give rise to the same physical solutions, the Hamilton-Jacobi method is physically more appealing. Often, the space of equilibrium states is the space of externally controllable parameters which control the equilibrium state of the system under consideration. The Hamilton-Jacobi approach allows for the description of thermodynamic transformations on the space of equilibrium states, often in terms of the externally controllable parameters and not the entire thermodynamic phase space. As in the case with classical mechanics, the dynamics of the remaining thermodynamic variables which do not lie on the space of equilibrium states can be simply obtained from the knowledge of the time evolving principal function. This is explicitly illustrated by considering simple examples. \\

With this motivation, we present the organization of this paper. The next section is introductory wherein the basic notions of contact geometry and its connection with thermodynamics are discussed very briefly with the aim of making this paper self contained. Following this, in section-(\ref{HJSection}), we describe the general form of the Hamilton-Jacobi equation relevant to studying thermodynamic transformations and make several comments on the choice of the principal function and the nature of the dynamics described in this formalism. In section-(\ref{Examplesection}), we work out some simple examples of thermodynamic transformations of the ideal gas and solve the problems completely using the Hamilton-Jacobi method without referring to the standard Hamiltonian procedure. Subsequently, we discuss the geometric Hamilton-Jacobi formulation in section-(\ref{GHJTS}), together with some examples. We end with some remarks in section-(\ref{discuss}).


\section{Contact geometry and thermodynamics} \label{contactBH}
We shall briefly review some basic aspects of contact geometry in the following subsection. The reader is referred to  \cite{arnoldcm, Geiges,Arnold,CM1,CM2} for a detailed introduction. In subsection-(\ref{connectthermo}), we describe its connection with equilibrium thermodynamics.

\subsection{Contact geometry} \label{contactG}
Contact geometry is the odd dimensional cousin of the more familiar symplectic geometry used to describe phase spaces in standard Hamiltonian dynamics. The central notion is that of a contact manifold which is defined to be a pair \((\mathcal{M},\eta)\) where \(\mathcal{M}\) is a smooth manifold of real dimension \(2n+1\) while \(\eta\) is a one-form satisfying the condition of complete Frobenius non-integrability: 
\begin{equation}\label{volume}
	\eta \wedge (d\eta)^n \neq 0.
\end{equation} This is equivalent to saying that the distribution generated as \({\rm ker}(\eta)\) is maximally non-integrable in the Frobenius sense. Here, \(\eta \wedge (d\eta)^n\) is the volume form on \(\mathcal{M}\). Now associated with \(\eta\) there exists a global vector field \(\xi\), known as the Reeb vector field uniquely defined through the relations
\begin{equation}\label{xidef}
	 \eta(\xi) = 1, \hspace{5mm} d\eta(\xi,\cdot) = 0.
\end{equation} 
One may think of \(\eta\) to be in a sense, dual to the vector field \(\xi\). Analogous to the symplectic case, there is a Darboux theorem for contact manifolds which asserts the existence of local coordinates \((s,q^i,p_i)\) in any open patch on \(\mathcal{M}\) such that
\begin{equation}
	\eta = ds - p_i dq^i, \hspace{5mm} \xi = \frac{\partial}{\partial s},
\end{equation} which satisfy the conditions given in Eqs. (\ref{volume}) and (\ref{xidef}). \\

Let us recall that on a symplectic manifold, there is a vector bundle isomorphism between the tangent and cotangent bundles due to the symplectic form which is both closed and non-degenerate. Such an isomorphism associates a vector field \(X_H\) to the one-form \(dH\) for a given Hamiltonian function \(H\). Analogously, in contact geometry, there exists a map \(h \mapsto X_h\) given by\footnote{In the published version: Pramana - J Phys. 97, 49 (2023), this map is erroneously indicated as a vector bundle isomorphism.}
\begin{equation}
	\eta(X_h) = - h, \hspace{5mm} d\eta(X_h, \cdot) = dh - \xi(h)\eta .
\end{equation}
Quite naturally, \(X_h\) is known as the contact Hamiltonian vector field associated with the contact Hamiltonian function \(h \in C^\infty(\mathcal{M})\) . It is simple to check that the local coordinate expression for \(X_h\) reads 
\begin{equation}\label{contactfield}
	X_h = \bigg(  p_i \frac{\partial h}{\partial p_i} - h\bigg) \frac{\partial}{\partial s}  + \bigg( \frac{\partial h}{\partial p_i} \bigg) \frac{\partial}{\partial q^i} - \bigg(  p_i \frac{\partial h}{\partial s} + \frac{\partial h}{\partial q^i}\bigg) \frac{\partial}{\partial p_i} .
\end{equation}

A particularly interesting feature from the point of view of thermodynamics is the existence of a special class of submanifolds (of a contact manifold) of maximal dimension whose tangent spaces are contained in the kernel of the contact form \(\eta\) at any point. In other words, they are solutions to the equation \(\eta=0\). Let \(L \subset \mathcal{M}\) be a submanifold of a contact manifold \((\mathcal{M},\eta)\) and \(f: L \rightarrow \mathcal{M}\) be the relevant inclusion map. If \(L\) be a maximal dimensional integral submanifold such that \(f^*\eta=0\), then \(L\) is called a Legendre submanifold. It can be shown that the maximal dimension is \(n\) and locally the general form of such a submanifold \(L\) is given by \cite{arnoldcm}
\begin{equation}
   p_i = \frac{\partial F}{\partial q^i}, \hspace{3mm} q^j = -\frac{\partial F}{\partial p_j},  \hspace{3mm} s = F - p_j \frac{\partial F}{\partial p_j}, \, 
\end{equation} where \(I \cup J\) is a disjoint partition of the set of indices \(\{1,2,....,n\}, i \in I, j \in J\) and \(F=F(q^i,p_j)\) is a function of \(n\) variables known as the generator of Legendre submanifold \(L\). This implies that not all \(n\) dimensional submanifolds of \(\mathcal{M}\) are Legendre submanifolds: coordinates on a Legendre submanifold cannot include a conjugate pair.\\

It should be remarked that contact Hamiltonian dynamics is such that a Legendre submanifold \(L\) is invariant to the flow of \(X_h\) if and only if the contact Hamiltonian function \(h\) vanishes on \(L\)~\cite{RT5}\footnote{We exclude the possibility that \(\xi(h) = 0\) since this case is not very interesting.}. This means that if \(X_h\) be the contact vector field generated by \(h\), such that \(h|_L=0\) where \(L\) is a particular Legendre submanifold, then, if the flow of \(X_h\) enters \(L\), it stays on \(L\). This follows from the fact that \(\dot{h}=0\) whenever \(h=0\) and hence \(X_h\) is tangent to the level surface for which \(h=0\). In other words, a contact vector field \(X_h\) is tangent to a Legendre submanifold \(L\) if and only if \(L \subset h^{-1}(0)\). With this background, we can make connection with thermodynamics.

\subsection{Connection with thermodynamics}\label{connectthermo}
We can now see how contact geometry naturally and elegantly describes classical thermodynamics. We define the thermodynamic phase space to be a contact manifold \((\mathcal{M},\eta)\). In terms of the Darboux coordinates, the contact one-form can be expressed as
\begin{equation}\label{contactstructure}
\eta = ds - p_1dq^1 - p_2dq^2 - \cdots  \, .
\end{equation}
Recall that for a reversible thermodynamic process, one has
\begin{equation}\label{firstlaw}
 dU - TdS + PdV = 0 \, .
\end{equation}
Now that thermodynamic systems at equilibrium satisfy Eq. (\ref{firstlaw}), we immediately identify them as Legendre submanifolds of the contact thermodynamic phase space. This means that from Eq. (\ref{contactstructure}) with \(q^1 = S, q^2 = V\) we get
\begin{equation}
  s = U(S,V), \hspace{3mm} p_1 = \frac{\partial U}{\partial S} = T, \hspace{3mm} p_2 = \frac{\partial U}{\partial V} = -P \, . \hspace{3mm}
\end{equation}
Here, the generator of the Legendre submanifold is the internal energy \(U = U(S,V)\). Referring to Eq. (\ref{contactstructure}) it is clear that we could have chosen a different representation of the system where some other thermodynamic potential would have the role of the generating function for the Legendre submanifold. These thermodynamic systems can be interpreted geometrically~\cite{RT1,RT2} as the triplet \((\mathcal{M},\eta,L)\) where \(L\) is the Legendre submanifold corresponding to that particular system in the thermodynamic phase space \((\mathcal{M},\eta)\). A contact vector field \(X_h\) generated by a contact Hamiltonian \(h\) can be considered as a generator of a thermodynamic process on a Legendre submanifold \(L\), if the Hamiltonian function vanishes on the appropriate submanifold, i.e. \(h|_L=0\). The flow of such a vector field is tangent to the equilibrium submanifold and the flow stays on \(L\). It means that if an initial point is taken to be on \(L\), i.e. it is an equilibrium state, all the subsequent points lie on \(L\) because \(h\) is conserved along the level surface \(h = 0\). Therefore, a thermodynamic system undergoing a particular transformation is interpreted as the quadruple \((\mathcal{M},\eta,h,L)\) where \(h=0\) on \(L\).

\section{Hamilton-Jacobi approach to thermodynamic transformations}\label{HJSection}
In this section, we introduce formally, a Hamilton-Jacobi theory whose characteristic curves correspond to thermodynamic transformations. We shall be following the contact Hamilton-Jacobi formalism presented in \cite{CM2} (see also \cite{CM1,RT6,deLeon,deLeon2,Wada} for some related works). 

\subsection{Concept of principal function}
In the Hamilton-Jacobi theory, the central notion is that of the principal function which is defined on the configuration space, i.e. it is only a function of one half of the conjugate variables excluding any conjugate pair. The analogous concept of configuration space in contact geometry is that of a Legendre submanifold. Thus, the principal function we seek should be defined on the Legendre submanifold representing the system. Let \(\{q^i\}\) be the local coordinates parametrizing the Legendre submanifold representing the thermodynamic system of interest and \(W = W(q^i,t)\) be the principal function. Then, the Hamilton-Jacobi formalism seeks a partial differential equation of the form (see \cite{CM2} for some relevant details):
\begin{equation}\label{calf}
	\mathcal{F}\Bigg(q^i,W,t, \frac{\partial W}{\partial q^i} , \frac{\partial W}{\partial t} \Bigg) = 0.
\end{equation}
If the thermodynamic transformation in the phase space is generated by the contact Hamiltonian function \(h\), then upon defining \(\mathcal{F} = E - h\), the solution to Eq. (\ref{calf}) on the configuration space turns out to be 
\begin{equation}
	\eta_E = dW - p_i dq^i + h dt = 0,
\end{equation} or equivalently, the following relations hold: 
\begin{equation}\label{hjeqn}
	p_i = \frac{\partial W(q^i,t)}{\partial q^i}, \hspace{7mm} \frac{\partial W(q^i,t)}{\partial t} + h \bigg( q^i, \frac{\partial W(q^i,t)}{\partial q^i} \bigg) = 0.
\end{equation}
Notice that \(\eta_E\) can be interpreted as an extended version of the Poincare-Cartan one-form encountered in classical mechanics. This is indeed defined on the configuration space, for we have replaced all the momenta by the derivatives of the principal function with respect to the corresponding coordinates. From thermodynamics, it is clear from the first among Eqs. (\ref{hjeqn}) that the thermodynamic potential, say the internal energy is a candidate for the principal function. The second equation, which has been called the contact Hamilton-Jacobi equation shall be verified through several examples in the next section.\\

Thus, we can formulate thermodynamic transformations described by a contact Hamiltonian solely on the Legendre submanifold of the system such that the characteristic curves of the Hamilton-Jacobi equation correspond to these transformations. Here \(\{q^i\}\) are the independent thermodynamic variables and \(W = W(q^i,t)\) is a dynamical fundamental relation for the system of interest, i.e. the other \(n\) thermodynamic quantities and their time evolution are obtained by taking derivatives of \(W(q^i,t)\).

\subsection{Statistical ensembles}
In the previous subsection, we described the notion of the principal function in the setting of Hamilton-Jacobi theory for a thermodynamic system or any contact Hamiltonian system. We shall now describe statistical ensembles and representations in this context. If we consider a hydrostatic system, then the first law of thermodynamics is given by
\begin{equation}\label{firstlawhydrostatic}
 dU - TdS + PdV - \mu dN = 0.
\end{equation} Comparing with the expression \(\eta = ds - p_i dq^i\), we identify \(s = U\), \((q^1,q^2,q^3) = (S,V,N)\), \((p_1,p_2,p_3) = (T,-P,\mu)\). The equilibrium condition [Eq. (\ref{firstlawhydrostatic})] implies that \(\eta = 0\) on the Legendre submanifold representing the system and \((S,V,N)\) are the independent thermodynamic variables. At equilibrium, one identifies \(W = U(S,V,N)\) as the principal function of the system and the other thermodynamic variables can be derived as derivatives of \(U(S,V,N)\) with respect to its arguments: 
\begin{widetext}
\begin{equation}\label{energyrepdef}
	T = \Bigg( \frac{\partial U(S,V,N)}{\partial S} \Bigg)_{V,N}, \hspace{5mm}  P = - \Bigg( \frac{\partial U(S,V,N)}{\partial V} \Bigg)_{S,N}, \hspace{5mm} \mu = \Bigg( \frac{\partial U(S,V,N)}{\partial N} \Bigg)_{S,V}. 
\end{equation}
\end{widetext}
 Note that this is a microcanonical description, for we can rearrange Eq. (\ref{firstlawhydrostatic}) to write
\begin{equation}\label{firstlawhydrostatic2}
dS - \frac{dU}{T} - \frac{P}{T}dV + \frac{\mu}{T} dN = 0,
\end{equation} where we may now identify as independent thermodynamic variables \((q^1,q^2,q^3) = (U,V,N)\) and the principal function \(W = S (U,V,N)\). As before, the other thermodynamic variables can be obtained as derivatives of the principal function, i.e. 
\begin{widetext}
 \begin{equation}
 	T = \Bigg(\frac{\partial S(U,V,N)}{\partial U}\Bigg)^{-1}_{V,N}, \hspace{5mm} P=  T\Bigg(\frac{\partial S(U,V,N)}{\partial V}\Bigg)_{U,N}, \hspace{5mm} \mu = -T \Bigg(\frac{\partial S(U,V,N)}{\partial N}\Bigg)_{U,V}. 
 \end{equation}
 \end{widetext}
 Therefore, in general, points on the Legendre submanifold representing the system can be parametrized by taking as independent variables any three among \(U\), \(S\), \(V\) and \(N\) while the other one can be taken to be the principal function. This leads to the notion of a representation of a thermodynamic system. While Eq. (\ref{firstlawhydrostatic}) is known as the \textit{energy representation}, Eq. (\ref{firstlawhydrostatic2}) is the \textit{entropy representation}. Both of them represent the same system in the microcanonical ensemble. Further, one can take the volume of the system to be the principal function by rearranging Eq. (\ref{firstlawhydrostatic}) to give the \textit{volume representation} as
 \begin{equation}
 	dV + \frac{dU}{P} - \frac{T}{P} dS - \frac{\mu}{P} dN = 0,
 \end{equation} where the independent thermodynamic variables are \((q^1,q^2,q^3) = (U,S,N)\). The other thermodynamic variables are given by
 \begin{widetext}
 \begin{equation}
 	P = -\Bigg(\frac{\partial V(U,S,N)}{\partial U}\Bigg)^{-1}_{S,N}, \hspace{5mm}  T= P \Bigg(\frac{\partial V(U,S,N)}{\partial S} \Bigg)_{U,N}, \hspace{5mm} \mu  = P \Bigg( \frac{\partial V(U,S,N)}{\partial N} \Bigg)_{U,S}.
 \end{equation}
 \end{widetext}
Similarly, we may consider \(N = N(U,S,V)\) to be the principal function by suitably rearranging Eq. (\ref{firstlawhydrostatic}) giving the \textit{number of particles representation}, i.e. 
 \begin{equation}
 	dN - \frac{dU}{\mu} + \frac{T}{\mu} dS - \frac{P}{\mu} dV = 0,
 \end{equation}
 \begin{widetext}
 or equivalently, 
  \begin{equation}
 	\mu = \Bigg(\frac{\partial N(U,S,V)}{\partial U}\Bigg)^{-1}_{S,V}, \hspace{5mm}  T= - \mu \Bigg(\frac{\partial N(U,S,V)}{\partial S} \Bigg)_{U,V}, \hspace{5mm} P  = \mu \Bigg( \frac{\partial N(U,S,V)}{\partial V} \Bigg)_{U,S}.
 \end{equation}
\end{widetext}

Therefore, in a given ensemble, the choice of the principal function is not unique. For the present case, which is the microcanonical ensemble, one may choose the principal function to be one amongst the thermodynamic variables \((U,S,V,N)\). Following arguments akin to those invoked above for the microcanonical ensemble, for a more general case one can make the following proposition: 

\begin{prop}\label{prop1}
	Consider a thermodynamic system whose variables \(\{q^i\}\) with \(i = 1, \cdots, n\)  are controlled by boundary conditions, i.e. either by external baths or boundaries. Let \(\Phi = \Phi(q^i)\) be a relevant potential function. On the Legendre submanifold representing the system, one can pick as the principal function, any one among the \(n+1\) variables \(\{\Phi,q^i\}\) such that it becomes a function of the other \(n\) variables obtained by solving the relation \(\Phi = \Phi(q^i)\). The remaining \(n\) conjugate thermodynamic variables characterizing the system can be obtained from the derivatives of the principal function so chosen. 
\end{prop}

We point out that although this is generically true, in practice for an arbitrary system, one might not be able to solve the equation \(\Phi = \Phi(q^i)\) for any arbitrary \(q^i\). This can be expected because thermodynamic potentials are complicated functions of their arguments. But nevertheless, proposition-(\ref{prop1}) stands as an appealing result arising due to the self consistent structure of equilibrium thermodynamics. Geometrically, if \(L\) is the Legendre submanifold representing the system, one can consider the space \(L \times \mathbb{R}\) which is locally \(\mathbb{R}^{n+1}\). Then the equilibrium condition which is equivalent to taking one variable amongst the set \(\{\Phi,q^i\}\) to be a function of the remaining \(n\) variables defines a suitable projection (locally) \(f: \mathbb{R}^{n+1} \rightarrow \mathbb{R}^n\) where \(\mathbb{R}^n\) is the space of equilibrium states of the system in a particular representation. The choice of the principal function therefore depends on the particular representation chosen. However, one can carefully choose the contact Hamiltonian function in both representations such that they are equivalent. \\

\begin{prop}\label{prop2}
Consider a thermodynamic system with potential \(\Phi = \Phi(q^i)\) which satisfies: \(d\Phi - p_i dq^i = 0\). Let us now have a different representation obtained by taking \(q^k = q^k (\Phi,q^1, \cdots, q^{k-1}, q^{k+1}, \cdots, q^n)\) where \(1 \leq k \leq n\) as the thermodynamic potential such that at equilibrium one has \(dq^k - (1/p_k) d\Phi + (p_1/p_k) dq^1 + \cdots = 0\). Then, if a given thermodynamic transformation in the \(\Phi\) representation is generated by the contact Hamiltonian \(h\), then the equivalent transformation in the \(q^k\) representation is described by the contact Hamiltonian \(-h/p_k\). 
\end{prop}

\textit{Proof -} Consider the \(\Phi\) representation with first law \(d\Phi - p_i dq^i = 0\). Now let there be a thermodynamic transformation due to the contact Hamiltonian \(h\).  Although at stationarity, \(\Phi\) does not carry any time dependence, implicit or explicit, under a thermodynamic transformation, the potential and its arguments are time varying in general, i.e. \(q^i = q^i(t)\) and \(\Phi = \Phi(q^1, \cdots, q^n, t)\). The Hamilton-Jacobi equation then reads
\begin{equation}
	\frac{\partial \Phi(q^1, \cdots, q^n, t)}{\partial t} + h = 0.
\end{equation} On the other hand, the total time derivative of \(\Phi\) is given by
\begin{equation}
	\dot{\Phi} = \frac{\partial \Phi}{\partial t} + \frac{\partial \Phi}{\partial q^i} \dot{q}^i = - h + p_i \dot{q}^i.
	\end{equation} If we divide by \(p_k\), the above equation can be rearranged to give
	\begin{equation}
		\dot{q}^k = \frac{h}{p_k} + \frac{\dot{\Phi}}{p_k} - \frac{p_2 \dot{q}^2}{p_k} - \cdots.
	\end{equation}
	Considering the \(q^k\) representation where one has \(dq^k - (1/p_k) d\Phi + (p_1/p_k) dq^1 + \cdots = 0\), the above equation can be expressed in terms of the new coordinates \(\{Q^j\}\) where \(Q^1 = \Phi\), \(Q^2 = q^1\), \(\cdots\), and new momenta \(\{P_j\}\) with \(P_1 = 1/p_k\), \(P_2 = -p_2/p_k\), \(\cdots\) as
	\begin{equation}
		\dot{q}^k = P_1 h + P_j \dot{Q}^j. 
	\end{equation} \\
	Now, since \(\partial q^k / \partial Q^j = P_j\), we must get
	\begin{equation}
		\dot{q}^k = P_1 h + \frac{\partial q^k}{\partial Q^j} \dot{Q}^j. 
	\end{equation} Thus, one must put \( -h/p_k\) to be the new contact Hamiltonian so that the Hamilton-Jacobi equation in the \(q^k\)-representation reads
	\begin{equation}
		\frac{\partial q^k (Q^j,t)}{\partial t} -\frac{h}{p_k} = 0.
	\end{equation}
	For example, if a certain contact Hamiltonian \(h\) generates a specific thermodynamic transformation in the energy representation, then the contact Hamiltonian \( - \beta h\) (\(\beta = 1/T\) and we will set \(k_B =1\) in subsequent discussions) generates an equivalent dynamics in the entropy representation. We shall verify this in the examples presented in section-(\ref{Examplesection}). \\

One should note that although one is using several coordinate parametrizations to describe the same system, the statistical ensemble stays the same, i.e. there is a fixed choice of boundary conditions. Therefore, a mere change of coordinate parametrization does not lead to a change in the ensemble, say from microcanonical to canonical. Shifting from one ensemble to another, i.e. changing the boundary conditions, requires Legendre transforms. We discuss them below.

 \subsubsection{Legendre transforms}
 
 In thermodynamics, one can switch between different ensembles by performing Legendre transforms. For example, in the canonical ensemble, the equilibrium state of a hydrostatic system is specified by the thermodynamic variables \((T,V,N)\). An appropriate function of these variables is the Helmholtz-Massieu potential obtained by the partial Legendre transform of the entropy: 
 \begin{equation}\label{legendre}
 	\Psi = -\frac{F}{T} =  S -  \frac{U}{T},
 \end{equation} thereby swapping \(U\) with \(T\), i.e.  \((U,V,N) \rightarrow (T,V,N) \). The first law of thermodynamics in the canonical ensemble reads
 \begin{equation} \label{Psifirstlaw}
 	d\Psi - \frac{U}{T^2} dT - \frac{P}{T}dV + \frac{\mu}{T} dN = 0.
 \end{equation}
 Thus, the other thermodynamic quantities are given by
 \begin{widetext}
  \begin{equation}
 	U= T^2 \Bigg(\frac{\partial \Psi(T,V,N)}{\partial T}\Bigg)_{V,N}, \hspace{5mm}  P= T \Bigg(\frac{\partial \Psi(T,V,N)}{\partial V}\Bigg)_{T,N}, \hspace{5mm} \mu  = -T \Bigg(\frac{\partial \Psi(T,V,N)}{\partial N}\Bigg)_{T,V}.
 \end{equation}
 \end{widetext} More generally, following proposition-(\ref{prop1}), one may consider as a principal function, any one among \((\Psi,T,V,N)\), which is then a function of the other three. These four cases are the equivalent parametrizations for a hydrostatic system in the canonical ensemble and describe the same Legendre submanifold. It turns out that the Legendre transform connecting the microcanonical and canonical descriptions [Eq. (\ref{legendre})] defines a different Legendre submanifold altogether. Usually, these two Legendre submanifolds, say \(L\) and \(L'\) are diffeomorphic to one another. This can be ensured if the Legendre transformation connecting the two is regular \cite{RT7}, i.e.  if a partial Legendre transform swaps the variables \(\{q^\alpha\}\) with \(\{p_\alpha\}\) for some value(s) of \(\alpha\) (or dummy index \(\beta\)) lying in the range \(1, \cdots, n\), the principal function of the first ensemble satisfies
 \begin{equation}\label{regularity}
 	\Bigg| \frac{\partial^2 W}{\partial q^\alpha \partial q^\beta} \Bigg| \neq 0.
 \end{equation}
On the other hand, in any case, if Eq. (\ref{regularity}) is not satisfied, the Legendre transform is singular leading to inconsistencies. We shall not refer to such situations here. \\

So far we have discussed the choice of the principal function for a thermodynamic system. Once a consistent choice has been made and a thermodynamic transformation takes place, the Hamilton-Jacobi equation describes completely the dynamical information. In the next section, we consider examples to illustrate this point.

\section{Thermodynamic transformations} \label{Examplesection}
In this section, we consider examples of thermodynamic transformations. Considering specific thermodynamic transformations generated by a contact Hamiltonian, we reformulate the problem as a Hamilton-Jacobi theory. It is found that the same equations of motion can be recovered using the latter method. Conceptually, this is more satisfying because the principal function depends on the set of parameters which are experimentally controlled. The conjugate thermodynamic variables assume a secondary role and can be derived from the knowledge of the dynamical principal function. For simplicity, we consider the example of the ideal gas. It should be noted that \(U = U(S,V,N)\) or \(S = S(U,V,N)\) are homogenous of degree one in each argument, i.e. \(U(\lambda S, \lambda V, \lambda N) = \lambda U (S,V,N)\) and similarly \(S(\lambda U, \lambda V, \lambda N) = \lambda S(U,V,N)\) where \(\lambda\) is some scale factor. Thus, Euler's theorem for homogenous functions tells us that \(U = TS - PV + \mu N\). This relationship, known as the Euler relation will be frequently referred to in the following examples. 

\subsection{Isochoric process of the ideal gas}\label{isochoricsubsection}
We consider an isochoric or constant volume transformation of the ideal gas. Both the energy and entropy representations are considered and their mutual consistency is shown. We begin with the simpler energy representation below.

\subsubsection{Energy representation}
In the energy representation, the relevant thermodynamic variables for describing the ideal gas are given as \(\{U,T,S,P,V,\mu,N\}\), in a seven dimensional thermodynamic phase space. Eq. (\ref{firstlawhydrostatic}) is identified with the vanishing of the contact form defining an equilibrium submanifold. Therefore, the conjugate variable pairs are: \((q^1,p_1) \rightarrow (S,T)\); \((q^2,p_2) \rightarrow (V,-P)\); \((q^3,p_3) \rightarrow (N,\mu)\) with \(s=U\). The internal energy \(U\) is the generating function for the appropriate Legendre submanifold which may be expressed by solving the Sackur-Tetrode equation \(S = S(U,V,N)\) to give \(U = U(S,V,N)\), which reads
\begin{equation}\label{internal energy}
  U(S,V,N) = A \exp[S/CN]V^{-1/C}N^{1+1/C}, \, 
\end{equation}
where \(C\) is the specific (per particle) heat capacity at constant volume and \(A > 0\) is an appropriate constant. We can easily recover the ideal gas equation \(PV=NT\) from Eq. (\ref{internal energy}) by taking first derivatives. To generate an isochoric process, consider the contact Hamiltonian of the form
\begin{equation}\label{hisochor}
	h = TS + \mu N - \gamma U,
\end{equation} where \(\gamma = (1+C)/C\). Note that on the Legendre submanifold representing the system, one has the relations \(PV=NT\) and \(U = CNT\), and consequently \(h = TS - NT +\mu N - U = 0\) due to the Euler relation for the ideal gas. Thus, as mentioned in subsection-(\ref{connectthermo}), the resulting dynamics corresponds to a thermodynamic transformation of the ideal gas. In other words, if an initial point on a dynamical trajectory is taken to lie on the Legendre submanifold representing the system, i.e. the initial point is an equilibrium state of the system, then all the subsequent points are equilibrium states and the trajectory lies on the Legendre submanifold. We would now formulate a Hamilton-Jacobi equation for dynamics described by this choice of the contact Hamiltonian. For that we first note that in this energy representation, the natural choice of the principal function is internal energy. It must satisfy the contact Hamilton-Jacobi equation meaning that we should have
\begin{equation}\label{HJ1}
	\frac{\partial U}{\partial t} + S \frac{\partial U}{\partial S} + N \frac{\partial U}{\partial N}  = \gamma U,
\end{equation} where \(U = U(S,V,N,t)\). We must make two general remarks before proceeding any further. First, if in a thermodynamic process, all the intermediate points are to be equilibrium points then the process must be quasi-static, i.e. it must proceed with an infinitesimally small speed such that at every instant, the system passes through an equilibrium state (except for the dynamics discussed in subsection-(\ref{OCS})). If this is kept in mind, the variable \(t\) appearing in the above equation is not really time but an affine parameter which parametrizes the integral curves of the thermodynamic process. In this manner, we avoid the notion of finite time transformations wherein the system may pass through intermediate steps which are not exactly equilibrium states. However, for the sake of brevity, we shall keep referring to \(t\) as time. Second, the internal energy (or the principal function) of the system carries both an implicit and an explicit time dependence in a general case. It is explicitly time dependent if its partial derivative with respect to \(t\) is non-zero. In addition, it may depend on time through its time evolving arguments: \(S\), \(V\) and \(N\). In our notation, we denote with a dot, a total time derivative, i.e. \(\dot{f} = df/dt\) while a partial time derivative is denoted by the operator \(\partial/\partial t\). \\

We shall now solve the Hamilton-Jacobi equation [Eq. (\ref{HJ1})] to obtain the time evolution of the principal function and its arguments. In order to do so, we first consider separation variables through which the explicit and implicit time dependences of \(U\) can be separated. Let us consider the ansatz, \(U(S,V,N,t) = F(t) G(S,V,N)\). Note that although \(\partial G/\partial t = 0\), it is expected to carry an implicit time dependence due to the time evolution of \(S\), \(V\) and \(N\), i.e. \(\dot{G} \neq 0 \). However, the implicit time dependence arises once the equations of motion are solved and therefore can be suppressed at this stage. In terms of \(F\) and \(G\), we have 
\begin{equation}
	\frac{dF}{dt} = \lambda F, \hspace{7mm} S \frac{\partial G}{\partial S} + N \frac{\partial G}{\partial N} + 0 \times \frac{\partial G}{\partial V}= (\gamma - \lambda) G,
\end{equation}
where \(\lambda\) is a separation constant. The first equation is an ordinary differential equation which can be integrated to give \(F(t) \sim e^{\lambda t}\). The second equation which is still a partial differential equation can be solved by the well known method of characteristics. On the solution surface, if \(t\) parametrizes the trajectories then the method of characteristics gives rise to the following ordinary differential equations with \(t\) as the independent variable: 
\begin{equation}\label{SVNeqns0}
	\frac{d S}{d t} = S, \hspace{5mm} \frac{d N}{d t} = N, \hspace{5mm}  \frac{d V}{d t} = 0, \hspace{5mm} \frac{d G}{d t} = (\gamma-\lambda) G. 
\end{equation} These are the equations of motion for \(S\), \(V\) and \(N\) on the space of equilibrium states. The last equation above is integrated to give the implicit time dependence of \(G\) as: \(G(t) \sim e^{(\gamma - \lambda)t}\). Therefore, the overall dependence on \(t\) of \(U\) is \(U(t) = F(t) G(t) \sim e^{\gamma t}\) whose total time derivative gives
\begin{equation}
	\frac{d U}{d t} = \gamma U.
\end{equation} The curves parametrized by \(t\) are the solutions to the Hamilton-Jacobi problem. Let us note that all this while, we have been restricted to the Legendre submanifold representing the equilibrium states of the system rather than the entire thermodynamic phase space whose full information was irrelevant. The exact functional form of \(G\) requires statistical mechanics input as given in Eq. (\ref{internal energy}) and cannot be determined from the knowledge of \(h\) alone. Thus \(G = G(t)\) represents only the implicit dependence on \(t\) due to \((S,V,N)\) depending on \(t\) in their evolution. This lets us fix the separation constant \(\lambda\). Upon substituting for \(S\), \(V\) and \(N\) into Eq. (\ref{internal energy}), one finds
\begin{equation}
	G(t) \sim \exp[S_0/CN_0]V_0^{-1/C}N_0^{1+1/C} e^{\big(1 + 1/C\big)t},
\end{equation} where \(S_0\), \(V_0\) and \(N_0\) are the corresponding quantities at \(t = 0\). Comparing this with \(G(t) \sim e^{(\gamma-\lambda)t}\) fixes the value of the separation constant to be \(\lambda = 0\). This means that the internal energy has no explicit time dependence and its entire time evolution can be attributed to the evolution of \(S\) and \(N\). Suppressing the implicit dependence  on \(t\) into the respective variables \((S,V,N)\), the final expression for \(U = U(S,V,N,t)\) reads (putting \(\lambda = 0\))
\begin{equation}\label{UUU}
	U(S,V,N,t) = A \exp[S/CN]V^{-1/C}N^{1+1/C},
\end{equation}
 which is the solution of the Hamilton-Jacobi equation. Alternatively, one could have extracted the entire \(t\) dependence to write\begin{equation}
	U(S_0,V_0,N_0,t) = A \exp[S_0/CN_0]V_0^{-1/C}N_0^{1+1/C} e^{\gamma t}.
\end{equation} We refrain from doing so because we intend to derive the dynamical values of the conjugate thermodynamic variables by taking derivatives of \(U\) with respect to \((S,V,N)\) and not their initial values \((S_0,V_0,N_0)\). From here, one can easily compute the time dependence of the conjugate thermodynamic variables \(T,P,\mu\) using simple thermodynamic definitions. 
 \begin{widetext} Explicitly one has
\begin{equation} \label{TevolvingU}
	T = \bigg(\frac{\partial U(S,V,N,t)}{\partial S}\bigg)_{V,N,t} = \frac{A}{C}  \exp[S/CN]V^{-1/C}N^{1/C} = T_0 e^{t/C} ,
\end{equation} where \(T_0 = \frac{A}{C}  \exp[S_0/CN_0]V_0^{-1/C}N_0^{1/C}\). Thus, noting that \(1/C = \gamma -1\), we get \(T = T_0 e^{(\gamma - 1)t}\). Similarly, one can compute the pressure as
\begin{equation} \label{PevolvingU}
	P = -\bigg(\frac{\partial U(S,V,N,t)}{\partial V}\bigg)_{S,N,t} = \frac{A}{CV}  \exp[S/CN]V^{-1/C}N^{1+1/C} = P_0 e^{\big(1 + 1/C\big)t} ,
\end{equation} 
\end{widetext}where \(P_0 = \frac{A}{CV_0}  \exp[S_0/CN_0]V_0^{-1/C}N_0^{1+1/C}\) thereby giving \(P = P_0 e^{\gamma t}\). One may similarly compute the chemical potential. It can be easily checked that Eqs. (\ref{TevolvingU}) and (\ref{PevolvingU}) together with Eq. (\ref{UUU}) are consistent with the ideal gas equation \(PV = NT\) and the equipartition theorem \(U =C N T\). 

\subsubsection{Entropy representation}
We shall now consider the entropy representation in which the entropy of the system assumes the role of the principal function. From Eq. (\ref{firstlawhydrostatic2}), we have the identifications \((q^1,p_1) \rightarrow (U,\beta)\); \((q^2,p_2) \rightarrow (V,\beta P)\); \((q^3,p_3) \rightarrow (N, - \beta \mu)\) where \(\beta = 1/T\). In what follows, we shall treat \(\beta\), \(\beta P\) and \(-\beta \mu\) to be independent variables. The entropy is given by the Sackur-Tetrode equation: 
\begin{equation}\label{SackurTetrode}
S = N \ln \Bigg [\frac{KVU^C}{N^{C+1}}\Bigg] + (C + 1)N,
\end{equation} where \(K > 0\) is a constant with appropriate dimensions. In order to generate an isochoric transformation, we consider the following contact Hamiltonian function:
\begin{equation}
 h = -S - \beta \mu N + \gamma \beta U.
\end{equation}
 This contact Hamiltonian vanishes on the Legendre submanifold representing the system as a consequence of the Euler relation. We now formulate a consistent Hamilton-Jacobi equation for this problem in the entropy representation. Using standard definitions introduced earlier, the Hamilton-Jacobi equation reads
\begin{equation}
	\frac{\partial S}{\partial t} + \gamma U \frac{\partial S}{\partial U} + N \frac{\partial S}{\partial N} = S,
\end{equation} where \(S = S(U,V,N,t)\). As before, we proceed to solve this equation by separation of variables thereby separating the explicitly time dependent part by taking the ansatz \(S(U,V,N,t) = F(t) G(U,V,N)\). Therefore, we have
\begin{equation}
	\frac{dF}{dt} = \lambda F, \hspace{7mm} \gamma U \frac{\partial G}{\partial U} + N \frac{\partial G}{\partial N} + 0 \times \frac{\partial G}{\partial V}  = (1 - \lambda) G. 
\end{equation} This leads to \(F(t) \sim e^{\lambda t}\) and the second equation as before can be solved by the method of characteristics thereby giving
\begin{equation}
	\frac{d U}{dt} = \gamma U, \hspace{5mm} \frac{d V}{dt} = 0, \hspace{5mm} \frac{d N}{dt} = N, \hspace{5mm} \frac{d G}{dt} = (1 - \lambda) G.
\end{equation}
The first three equations agree with those obtained in the energy representation. Integrating the last one gives, \(G(t) \sim e^{(1 - \lambda) t}\). This gives the total time dependence of the entropy to be \(S = F(t) G(t) \sim e^t\) thereby giving \(\dot{S} = S\) consistently. With all this, Eq. (\ref{SackurTetrode}) becomes
\begin{equation}
 G(t) = \Bigg\{N_0 \ln \Bigg[\frac{KV_0U_0^C}{N_0^{C+1}}\Bigg] + (C + 1)N_0\Bigg\} e^t,
\end{equation} and comparing this with \(G(t) \sim e^{(1 - \lambda)t}\) fixes the value of the separation constant \(\lambda = 0\). Therefore, the solution to the Hamilton-Jacobi equation which is of the form \(S = S(U,V,N,t)\) is given by
\begin{equation}
 S(U,V,N,t) = N \Bigg\{ \ln \Bigg[\frac{KVU^C}{N^{C+1}}\Bigg] + (C + 1)\Bigg\},
\end{equation} where \(U\) and \(N\) are time evolving. The dynamics of the remaining thermodynamic quantities can be derived easily from standard definitions. For example, \(\beta\) reads
\begin{equation}
	\beta =  \bigg(\frac{\partial S(U,V,N,t)}{\partial U}\bigg)_{V,N,t} = \frac{NC}{U}.
\end{equation} Substituting \(N = N_0 e^t\) and \(U = U_0 e^{\gamma t}\) one finds
\begin{equation}
	\beta =  \beta_0 e^{(1 - \gamma)t},
\end{equation} with \(\beta_0 = N_0 C/U_0\). Similarly, one can determine the time evolution of variables \(\beta P\) and \(-\beta \mu\). Thus, the Hamilton-Jacobi formulations in the energy and entropy representations for the same dynamical problem are equivalent to each other. Its consistency with the standard Hamiltonian technique on contact manifolds can be verified by explicit calculation of \(X_h\) in both representations.

\subsection{Isochoric-isothermal process of the ideal gas}
As yet another example, we consider an isochoric-isothermal transformation of the ideal gas \cite{RT5} in the energy representation. Consider the contact Hamiltonian \(h = TS - NT +\mu N - U\). As in the previous example, the contact Hamiltonian function can be seen to vanish on the Legendre submanifold representing the ideal gas due to the Euler relation. Let us now reformulate this transformation as a Hamilton-Jacobi problem. For the given choice of \(h\), one can write the Hamilton-Jacobi equation to be
\begin{equation}
	\frac{\partial U}{\partial t} + (S-N) \frac{\partial U}{\partial S} + N \frac{\partial U}{\partial N}  = U,
\end{equation}
where \(U = U(S,V,N,t)\). Note the explicit dependence of \(U\) on \(t\). As in the previous example, the variables \(S\) and \(N\) shall be evolving with \(t\) and therefore, it shall result in an overall implicit dependence of \(U\) on \(t\) in addition to the explicit dependence it already has. In order to solve this equation, consider the separation of variables \(U(S,V,N,t) = F(t) G(S,V,N)\) where \(F\) and \(G\) are suitable well behaved functions of their arguments. This results in the following two differential equations
\begin{equation}
	\frac{dF}{dt} = \lambda F, \hspace{7mm} (S-N) \frac{\partial G}{\partial S} + N \frac{\partial G}{\partial N} + 0 \times \frac{\partial G}{\partial V}= (1 - \lambda) G ,
\end{equation} where \(\lambda\) is the separation constant. This gives \(F(t) \sim e^{\lambda t}\) and the method of characteristics gives the following ordinary differential equations:
\begin{equation}\label{SVNeqns}
	\frac{d S}{d t} = S - N, \hspace{5mm} \frac{d N}{d t} = N, \hspace{5mm}  \frac{d V}{d t} = 0, \hspace{5mm} \frac{d G}{d t} = (1-\lambda) G. 
\end{equation} The first three are the equations of motion for \(S\), \(V\) and \(N\). The equation for \(G\) can be solved to give \(G(t) \sim e^{(1-\lambda)t}\). Therefore, the overall dependence of \(U\) on \(t\) is \(U(t) = F(t) G(t) \sim e^t\) whose total time derivative gives \(\dot{U} = U\).\\

Upon substituting for \(S\), \(V\) and \(N\) into Eq. (\ref{internal energy}), one finds
\begin{equation}
	G(t) \sim \exp[S_0/CN_0]V_0^{-1/C}N_0^{1+1/C} e^{\big(1 + 1/C\big)t},
\end{equation} where \(S_0\), \(V_0\) and \(N_0\) are the corresponding quantities at \(t = 0\). Comparing this with \(G(t) \sim e^{(1-\lambda)t}\) fixes the value of the separation constant to be \(\lambda = -1/C\). Thus, one has the following solution to the Hamilton-Jacobi equation in the form \(U = U(S,V,N,t)\):
\begin{equation}
	U(S,V,N,t) = A \exp[S/CN]V^{-1/C}N^{1+1/C} e^{-t/C}.
\end{equation}
From here, one can easily compute the time dependence of the conjugate thermodynamic variables \(T,P,\mu\) using simple thermodynamic definitions. 
\begin{widetext}
For example, the temperature is
\begin{equation} 
	T = \bigg(\frac{\partial U(S,V,N,t)}{\partial S}\bigg)_{V,N,t} = \frac{A}{C}  \exp[S/CN]V^{-1/C}N^{1/C} e^{-t/C} = T_0,
\end{equation} where \(T_0 = \frac{A}{C}  \exp[S_0/CN_0]V_0^{-1/C}N_0^{1/C}\) and is independent of time. Thus the transformation is indeed an isochoric-isothermal one. Similarly, one can obtain the pressure as
\begin{equation}
	P =  -\bigg(\frac{\partial U(S,V,N,t)}{\partial V}\bigg)_{S,N,t} = \frac{A}{CV} \exp[S/CN]V^{-1/C}N^{1+1/C} e^{-t/C} = P_0 e^t.
\end{equation} 
where \(P_0 = \frac{A}{CV_0} \exp[S_0/CN_0]V_0^{-1/C}N_0^{1+1/C} \). \end{widetext}
 A similar calculation can be carried out for the chemical potential. It is simple to check that these equations consistently reproduce \(PV = NT\) and \(U = C N T\), i.e. the ideal gas equation of state and the equipartition theorem. Therefore, we formulated and completely solved the dynamics of an isochoric-isothermal process of the ideal gas on the Legendre submanifold representing the system.  Working in a similar manner, one can reformulate the same problem in the entropy representation and obtain the same physical solutions. Here we do not pursue it further. 
 
\subsection{Ideal gas \(\rightarrow\) interacting gas}
Here we discuss Hamiltonian flows under which equilibrium states of an ideal gas flow into those of an interacting gas with two-particle interactions. For this, consider the contact Hamiltonian given by \(h = a/V\) where \(a > 0\) is a constant. Note that here \( h \neq 0\) on the Legendre submanifold representing the system and consequently the flow of \(X_h\) is not tangent to it. Thus, if the initial point (\(t = 0\)) is chosen to be an equilibrium state of the ideal gas, i.e. lying on the Legendre submanifold representing the ideal gas, subsequent points are no longer equilibrium states of the ideal gas. In fact, as we shall see, the subsequent points on the integral curves are equilibrium states of an interacting gas with two-body interactions with strength characterized by the constant \(at\). The resulting dynamics therefore generates a family of interacting gases. Each constant slice of \(t\) represents a different system, i.e. an interacting gas with strength of interaction summarized by \(at\). The reader is referred to \cite{RT5,contactBH} for more details on such flows and transformations. \\

For the present case, in the energy representation, the Hamilton-Jacobi equation reads
\begin{equation}
	\frac{\partial U}{\partial t} + \frac{a}{V} =  0,
\end{equation} or equivalently,
\begin{equation}\label{UHJ}
 \frac{\partial U}{\partial t} + 0 \times \frac{\partial U}{\partial S} + 0 \times \frac{\partial U}{\partial V}  + 0 \times \frac{\partial U}{\partial N}  +\frac{a}{V} = 0,
\end{equation}where \(U = U(S,V,N,t)\). It therefore follows that the characteristic curves are \(\dot{S} = \dot{V} = \dot{N} = 0\). Thus, the time dependence of \(U\) is only explicit and the above equation can be integrated to give
\begin{equation}\label{UU}
U(S,V,N) = U_0(S_0,V_0,N_0) - \frac{at}{V},
\end{equation} where \(U_0\) is given by Eq. (\ref{internal energy}). Taking derivative of the above equation with respect to \(V\), one finds the evolution of pressure as
\begin{equation}\label{PP}
P = P_0 - \frac{at}{V^2},
\end{equation} where \(P_0 = - (\partial U_0/\partial V)_{S,N,t}\) with \(V = V_0\). Temperature and chemical potential can be found by simple definitions and turn out to be conserved along the flow. Thus, the dynamical flow due to the choice of contact Hamiltonian \( h = a/V\) generates a family of interacting gases with equations of state given by Eq. (\ref{PP}). \\

Alternatively, one may consider an entropy representation where \(S =S (U,V,N,t)\) is the principal function. In this case, one has to consider the Hamiltonian \(h = - \beta a /V\). Then the Hamilton-Jacobi equation reads
\begin{equation}
\frac{\partial S}{\partial t} - \frac{a}{V} \frac{\partial S}{\partial U} = 0.
\end{equation} If one takes the ansatz \(S(U,N,V,t) = F(t) G(U,V,N)\), then separating the explicit dependence on \(t\), one has \(F(t) \sim e^{\lambda t}\) and the following equations by the method of characteristics: 
\begin{equation}
	\frac{dU}{dt} = -\frac{a}{V}, \hspace{5mm} \frac{dV}{dt} = 0, \hspace{5mm} \frac{dN}{dt} = 0, \hspace{5mm} \frac{dG}{dt} = -\lambda 	G.
\end{equation} Thus, we obtain Eq. (\ref{UU}) by integrating the first equation above while the last equation gives \(G(t) \sim e^{-\lambda t}\) meaning that the total time dependence of entropy is \(S = F(t) G(t) \sim e^{\lambda t} e^{-\lambda t}\) or \(\dot{S} = 0\) as expected. Therefore, the energy and entropy representations both provide an equivalent description of the dynamics. 

\section{Geometric Hamilton-Jacobi formulation}\label{GHJTS}
For the sake of completeness, we describe the geometric Hamilton-Jacobi theory which has been developed in the recent times \cite{deLeon,deLeon2}. Since the space of equilibrium states \(L\) is a Legendre submanifold (with independent local coordinates \(\{q^i\}\)) of a contact manifold, the thermodynamic phase space \(\mathcal{M}\), one may think about the latter as being \(\mathcal{M} \approxeq T^*L \times \mathbb{R}\) with some contact form \(\eta\). \\

Consider a section of the canonical projection \(\pi: T^*L \times \mathbb{R} \rightarrow L\), being denoted as \(\sigma\), i.e. \(\sigma: L \rightarrow T^*L \times \mathbb{R}\). In local coordinates, \(q^i \mapsto \sigma (q^i) = (q^i, \sigma_j (q^i), \sigma_s (q^i))\) where \(j = 1, 2, \cdots, n\). For a given \(h \in C^\infty (\mathcal{M})\), \(X_h\) is a vector field in \(T\mathcal{M} \approxeq T ( T^* L \times \mathbb{R})\).  Let us define the vector field \(X_h^\sigma = T\pi \circ X_h \circ \sigma\) on \(TL\). The above construction is summarized below.\\
\begin{center}
 \begin{tikzcd}[row sep=huge,column sep=huge]
T^*L \times \mathbb{R} \arrow[r, "X_h"] \arrow[d, "\pi"]
& T(T^*L \times \mathbb{R}) \arrow[d, "T\pi"] \\
L \arrow[r,  "X_h^\sigma"] \arrow[u, bend left,  "\sigma"]
& TL
\end{tikzcd}
\end{center}
For consistency, we require \(X_h \circ \sigma = T\sigma \circ X_h^\sigma\). This leads to the following conditions \cite{deLeon,deLeon2}: 
\begin{equation}\label{lll}
d(h \circ \sigma) = 0,
\end{equation} and, 
\begin{equation}
\sigma_j(q^i) = \frac{\partial \sigma_s (q^i)}{\partial q^j} ,
\end{equation}
where we have assumed that \(L\) is a Legendre submanifold, for which it can be shown \cite{deLeon2} that \(\sigma = j^1 \sigma_s\), i.e. it is the 1-jet of a function (which we take to be \(\sigma_s(q^i): L \rightarrow \mathbb{R}\)). Thus \(\sigma_s(q^i)\) plays the role of the principal function. For a section \(\sigma\) (of the canonical projection \(\pi\)) that satisfies these requirements, Eq. (\ref{lll}) or equivalently \(h \circ \sigma = k\) for some real constant \(k\) has been termed the geometric Hamilton-Jacobi equation. For our applications to thermodynamics, we shall choose \(k=0\) which implies
\begin{equation}\label{HJTG}
h \circ \sigma = 0. 
\end{equation} Let us apply this to two distinct examples. 

\subsection{Isochoric process of the ideal gas}
This particular example was discussed in the previous section and the contact Hamiltonian function in the energy representation is given by Eq. (\ref{hisochor}), i.e. \(h = TS + \mu N - \gamma U\). Here the coordinates on \(L\) are \(q^1 = S\), \(q^2 = V\) and \(q^3 = N\). Now the condition \(h \circ \sigma = 0\) gives
\begin{equation}\label{geomHJT}
\sigma_1(q^i) q^1 + \sigma_3(q^i) q^3 - \gamma \sigma_s (q^i)= 0,
\end{equation} or equivalently, 
\begin{equation}\label{geomHJT1}
 S \frac{\partial \sigma_s}{\partial S} +  N \frac{\partial \sigma_s}{\partial N} - \gamma \sigma_s = 0.
\end{equation}
It is simple to check by explicit differentiation that \(\sigma_s(q^i) = U(S,V,N)\) given in Eq. (\ref{internal energy}) satisfies Eq. (\ref{geomHJT1}) (because of Euler relation) and therefore is a solution to the geometric Hamilton-Jacobi equation. Here the variable \(V\) is in some sense, a cyclic coordinate. In the present case however, the form of \(U = U(S,V,N)\) is known from statistical mechanics and we just verified that it is consistent with the geometric Hamilton-Jacobi formulation. One can similarly consider the entropy representation and verify that \(S = S (U,V,N)\) given by the Sackur-Tetrode equation [Eq. (\ref{SackurTetrode})] satisfies the corresponding geometric Hamilton-Jacobi equation. 

\subsection{Onsager-Casimir dynamics}\label{OCS}
Let us consider a very different example, involving the general equation for nonequilibrium reversible-irreversible coupling (GENERIC) \cite{generic1,generic2,generic3,generic4,generic5} (see also \cite{generic6}). In a general setting, the GENERIC describes a time evolution of the form
\begin{equation}
\dot{q}^i = J^{ij} \frac{\partial E}{\partial q^j} + M^{ij} \frac{\partial S}{\partial q^j},
\end{equation} where \(J\) is an anti-symmetric matrix describing the reversible part of the evolution, whereas \(M\) is a positive symmetric matrix describing the irreversible part of the evolution \cite{generic1,generic6}. From a thermodynamic viewpoint, \(E = E(q^i)\) is the energy whereas \(S = S(q^i)\) is the entropy. Further, the following two degeneracy conditions are fulfilled \cite{generic1}: 
\begin{equation}\label{dc}
J^{ij} \frac{\partial S}{\partial q^j} =  M^{ij} \frac{\partial E}{\partial q^j} = 0. 
\end{equation}
Then it simply follows that
\begin{equation}
\frac{dE}{dt} = 0, \hspace{6mm} \frac{dS}{dt} = M_{ij} \frac{\partial S}{\partial q^i} \frac{\partial S}{\partial q^j} > 0,
\end{equation} which means although energy is conserved, entropy increases consistently due to the irreversible part of the evolution. Geometrically, this dynamics can be described using contact Hamiltonian dynamics \cite{generic1,generic3,generic4}. \\

Let us introduce the potential function \(\Phi (q^i) = - S(q^i) + \beta E(q^i)\). Then the degeneracy conditions [Eqs. (\ref{dc})] allow us to write
\begin{equation}\label{ocd}
\dot{q}^i = \beta^{-1} J^{ij} \frac{\partial \Phi}{\partial q^j} - M^{ij} \frac{\partial \Phi}{\partial q^j}.
\end{equation} This is known as the (nonlinear) Onsager-Casimir dynamics \cite{generic1}. Close to equilibrium, these equations can be linearized. Now the space of parameters \(q^i\) (denoted as \(L\)) can be lifted to a full thermodynamic phase space \(\mathcal{M} \approxeq T^*L \times \mathbb{R}\) as \(q^i \mapsto \sigma(q^i) = (q^i, \sigma_j (q^i), \sigma_s (q^i))\), where \(\sigma: L \rightarrow T^*L \times \mathbb{R}\) is a section of the canonical projection \(\pi: T^*L \times \mathbb{R} \rightarrow L\). \\

The following choice of contact Hamiltonian describes the Onsager-Casimir dynamics \cite{generic3,generic4}:
\begin{equation}
h = - \frac{1}{2} p_i M^{ij} p_j + \frac{1}{2} \frac{\partial \Phi}{\partial q^i} M^{ij} \frac{\partial \Phi}{\partial q^j} + \beta^{-1} p_i J^{ij}  \frac{\partial \Phi}{\partial q^j},
\end{equation}
in the sense that the corresponding contact vector field \(X_h\), upon being projected to \(L\) gives back Eq. (\ref{ocd}). Furthermore, it is easy to check that \(L\) is an invariant under the flow of \(X_h\), i.e. the vector field \(X_h\) is tangent to \(L\) on \(L\). So in the present case, the geometric condition Eq. (\ref{HJTG}) leads to the equation
\begin{equation}\label{HJTGeom3}
- \frac{1}{2} \sigma_i M^{ij} \sigma_j + \frac{1}{2} \frac{\partial \sigma_s}{\partial q^i} M^{ij} \frac{\partial \sigma_s}{\partial q^j} + \beta^{-1} \sigma_i J^{ij}  \frac{\partial \sigma_s}{\partial q^j} = 0,
\end{equation}
where we used the fact that \(\sigma_s (q^i) = \Phi(q^i)\) on \(L\). Due to the invariant nature of \(L\) under \(X_h\), Eq. (\ref{HJTGeom3}) is automatically satisfied and therefore, \(\sigma_s (q^i) = \Phi(q^i)\) is a solution of the geometric Hamilton-Jacobi equation [Eq. (\ref{HJTG})]. Thus, the vector fields \(X_h\) and \(X_h^\sigma\) are compatible in the sense that  \(X_h \circ \sigma = T\sigma \circ X_h^\sigma\). 

\section{Remarks}\label{discuss}
In this paper, we have formulated a consistent Hamilton-Jacobi approach for describing thermodynamic transformations. The problem is formulated and is solvable on the Legendre submanifold or space of equilibrium states representing the system of interest. The remaining thermodynamic variables which do not lie on the space of equilibrium states assume a secondary role and their time evolutions can be computed once the solution of the Hamilton-Jacobi problem is known. It has been explicitly verified in subsection-(\ref{isochoricsubsection}), that the energy and entropy representations provide an equivalent description of the dynamics in accordance with proposition-(\ref{prop2}). Note that although apriori, a contact Hamiltonian function is a function of all the phase space variables in general, within the Hamilton-Jacobi formalism, it is defined on the space of equilibrium states and is of the form \(h = h(q^i, W, \partial W/\partial q^i)\). \\

As it turns out, the equilibrium state of a thermodynamic system is characterized by the values of its externally controllable parameters, for example \((U,V,N)\) in the microcanonical ensemble or \((T,V,N)\) in the canonical ensemble. Thus, at least mathematically, one can always consider a representation where any of these variables are the independent ones. For the microcanonical ensemble, the entropy representation is the one where the externally controllable variables \((U,V,N)\) are taken to be independent and any thermodynamic transformation (which may experimentally be brought about by changing these variables) can be studied in terms of their time evolution in the Hamilton-Jacobi approach. It follows from the self consistent structure of thermodynamics that the conjugate variables which are \((T,P,\mu)\) can be described from the knowledge of the time evolving potential and its dynamical arguments. \\

There are several avenues for future work. For instance, the distance in the sense of thermodynamic length \cite{crooks1} between thermodynamic states related by a Hamiltonian transformation remains to be well understood. One can go further and consider defining an infinitesimal distance between the ideal gas and an interacting gas which are related by a Hamiltonian transformation such that the initial point is always an equilibrium state of the ideal gas whereas, subsequent points are not. To the best of our knowledge, this has not been studied before in detail. Furthermore, since the Hamilton-Jacobi approach completely describes dynamics on the space of externally controllable parameters, its relationship with optimal paths and optimal control protocols needs to be further investigated, especially in the context of molecular machines and microscopic engines \cite{crooks2,crooks3}. Such developments are expected to strengthen the connections between rigorous Hamiltonian theory, Riemannian geometry and thermodynamics. 

\section*{Acknowledgements}
The author is grateful to Chandrasekhar Bhamidipati for useful discussions. The financial assistance received from the Ministry of Education (MoE), Government of India, in the form of a Prime Minister's Research Fellowship (PMRF ID: 1200454) is gratefully acknowledged.

\end{document}